\pdfoutput=1
\documentclass{article}

\usepackage{jheppub}

\usepackage[utf8]{inputenc}

\usepackage{hyperref}

\usepackage{amsmath}
\usepackage{amssymb}
\usepackage{xcolor}
\usepackage{physics}
\usepackage{graphicx} 

\usepackage{pgf} 

\newcommand{\ads}{\mathrm{AdS}}
\newcommand{\ess}{\mathbb{S}}
\newcommand{\Vol}{\mathrm{Vol}}
\newcommand{\adsts}{{\ads \times \ess}}

\numberwithin{equation}{section}

\newenvironment{eqaed}
    {\begin{equation}
    \begin{aligned}
    }
    { 
    \end{aligned}
    \end{equation}
    }

\begin{document}

\title{Brane annihilation in non-supersymmetric strings}
\author{Riccardo Antonelli}
\emailAdd{riccardo.antonelli@sns.it}
\author{and Ivano Basile}
\emailAdd{ivano.basile@sns.it}
\affiliation{\emph{Scuola Normale Superiore and I.N.F.N.}\\\emph{Piazza dei Cavalieri 7, 56126, Pisa, Italy}}

\abstract{In this paper we study non-perturbative instabilities in Anti de-Sitter vacua arising from flux compactifications of string models with broken supersymmetry. In the semi-classical limit, these processes drive the vacua towards lower fluxes, which translate into higher curvatures and higher string couplings. In order to shed some light on this regime, we provide evidence for a description in terms of branes, which generate near-horizon $\ads$ throats. To this end, we study the attractor properties of the geometries near the throat, and we also characterize their asymptotics away from it. We also describe the instability within a probe-brane picture, finding an agreement between low-energy (super)gravity and brane instanton estimates of the decay rates.}

\maketitle


\section{Introduction}\label{sec:introduction}
    
    The issue of supersymmetry breaking is vital for string theory, for both theoretical and phenomenological reasons. A variety of mechanisms have been investigated, but they are all fraught with conceptual and technical obstacles, and primarily with the generic presence of instabilities. These may appear as tachyons, but even in tachyon-free models dilaton potentials arise due to Neveu-Schwarz (NS) tadpoles or quantum corrections to the vacuum energy. In both cases string theory back-reacts dramatically\footnote{In principle, one could address these phenomena by systematic vacuum redefinitions~\cite{Fischler:1986ci,Dudas:2004nd,Kitazawa:2008hv,Pius:2014gza}, but carrying out the program at high orders appears prohibitive.} on the original Minkowski vacua, whose detailed fate appears, at present, largely out of computational control.
    
    In this paper we build upon the results in~\cite{Mourad:2016xbk}, considering Anti de-Sitter ($\ads$) vacua in ten-dimensional tachyon-free string models~\cite{AlvarezGaume:1986jb,Sagnotti:1995ga,Sagnotti:1996qj,Sugimoto:1999tx} where, initially at least, large fluxes result in small string couplings and large $\ads$ radii, justifying the recourse to the low-energy (super)gravity. Specifically, we focus on the non-supersymmetric $SO(16) \times SO(16)$ heterotic model~\cite{AlvarezGaume:1986jb,Dixon:1986iz}, whose first quantum correction generates a dilaton potential, and on two orientifold models, the non-supersymmetric $U(32)$ ``Type $0'$B'' model~\cite{Sagnotti:1995ga,Sagnotti:1996qj} and the $USp(32)$ model~\cite{Sugimoto:1999tx} with ``Brane Supersymmetry Breaking'' (BSB)~\cite{Antoniadis:1999xk,Angelantonj:1999jh,Aldazabal:1999jr,Angelantonj:1999ms}, where a similar potential reflects the tension unbalance present in the vacuum. BSB is a particularly interesting phenomenon, since it combines a closed-string sector where supersymmetry is exact to lowest order with an open-string sector where supersymmetry is non-linearly realized\footnote{The original works can be found in~\cite{Sagnotti:1987tw,Pradisi:1988xd,Horava:1989vt,Horava:1989ga,Bianchi:1990yu,Bianchi:1990tb,Bianchi:1991eu,Sagnotti:1992qw}. For reviews, see~\cite{Dudas:2000bn,Angelantonj:2002ct,Mourad:2017rrl}.}~\cite{Dudas:2000nv,Pradisi:2001yv,Kitazawa:2018zys}. On the phenomenological side, the heterotic model has recently sparked some interest in non-supersymmetric model building~\cite{Abel:2015oxa,Blaszczyk:2015zta} while, in cosmological contexts, the peculiar behaviour of BSB~\cite{Sagnotti:2015asa,Gruppuso:2015xqa,Gruppuso:2017nap,Mourad:2017rrl} appears to provide a rationale for the low-$\ell$ lack of power in the Cosmic Microwave Background.
    
    These models feature $\adsts$ solutions, which are entirely specified by a flux number $n$, and large fluxes translate into parametrically small string couplings and curvatures for both $\ads$ and the internal sphere. The perturbative stability of $\ads_3 \times \ess^7$ and $\ads_7 \times \ess^3$ vacua of this type has been studied in~\cite{Gubser:2001zr,Basile:2018irz}, where unstable field modes were found to be present for low angular momenta. In the heterotic example, an antipodal projection suffices to eliminate them, while a more complicated projection, or a different choice of internal space, would be needed for the orientifolds. At any rate, here we shall find evidence that, in addition, non-perturbative (tunneling) instabilities tend to drive these vacua towards stronger couplings and higher curvatures. 
    
    In this work, we address in detail these non-perturbative instabilities, which manifest themselves as vacuum bubbles at the semi-classical level, and we compute the corresponding decay rates. We find that this tunneling process reduces the flux number $n$, thus driving the vacua away from the perturbative regime, albeit at a rate that is exponentially suppressed in $n$.
    
    We also recast these effects in terms of branes, drawing upon the analogy with the supersymmetric case where BPS brane stacks generate supersymmetric $\adsts$ near-horizon geometries. While Neveu-Schwarz (NS) five-branes in the heterotic $SO(16) \times SO(16)$ model of~\cite{AlvarezGaume:1986jb} appear more difficult to deal with in this respect, in orientifold models $\text{D}$-brane stacks provide a natural canditate for a microscopic description of these flux vacua and of their instabilities. Indeed, non-supersymmetric analogues of $\ads_5 \times \ess^5$ vacua in Type 0 strings, where tachyon condensation breaks conformal invariance of the dual gauge theory, were already described in terms of $\text{D}3$-branes in~\cite{Klebanov:1998yya}. In the non-tachyonic $0'\text{B}$ orientifold this role is played by the dilaton potential, which generates a running of the gauge coupling~\cite{Angelantonj:2000kh,Angelantonj:1999qg,Dudas:2000sn}. As a result, the near-horizon geometry is modified, and one recovers $\ads_5 \times \ess^5$ only in the limit of infinitely many $\text{D}3$-branes, when the supersymmetry-breaking dilaton potential becomes negligible.
    In contrast, $\text{D}1$ and $\text{NS}5$-branes should underlie the $\ads_3 \times \ess_7$ and $\ads_7\times \ess_3$ solutions found in~\cite{Mourad:2016xbk}. This might appear somewhat surprising, since $\text{D}p$-brane stacks in Type II string theory do not exhibit near-horizon geometries of this type for $p \neq 3$, but dress them instead with singular warp factors. Correspondingly, the dual gauge theory is non-conformal. While the emergence of a conformal dual involving $\text{D}1$ and $\text{NS}5$-branes in non-supersymmetric cases would be an enticing scenario, it is first necessary to establish whether brane descriptions of the $\adsts$ solutions to these models hold ground. In this paper we provide some evidence to this effect, studying the geometry generated by the branes at the level of the low-energy equations of motion. Specifically, we identify an attractor-like behaviour in the geometry near the $\ads$ throat that mirrors what happens in supersymmetric black holes, and we examine the asymptotics away from it. We then turn to the study of brane probes which, as we shall see, behave as suggested by the flux tunneling instability. In particular, we compute the corresponding decay rate in terms of the (super)gravity vacuum bubbles and also in terms of a brane instanton, obtaining consistent results. The brane picture that we propose can potentially open a computational window beyond the semi-classical regime, perhaps providing also a simpler realization of $\ads_3$/CFT$_2$ duality\footnote{The alternative case of $\ads_7$ could be studied, in principle, via $\text{M}5$-brane stacks.}. Moreover, in principle one could investigate these non-perturbative instabilities recasting them as holographic Renormalization Group (RG) flows in the dual gauge theory~\cite{Antonelli:2018qwz}.
    
    The paper is structured as follows. In Section~\ref{sec:ads_vacua} we describe in detail the low-energy EFT and the corresponding $\adsts$ solutions, which include a perturbative corner where both curvatures and the string coupling are parametrically small for large fluxes. In Section~\ref{sec:flux_tunneling} we study the flux tunneling process, and we present the computation of the semi-classical decay rate within the low-energy description. In Section~\ref{sec:brane_picture} we develop the microscopic picture, studying probe $\text{D}1$-branes and $\text{NS}5$-branes in the $\ads$ throat and matching their behaviour to the processes described in Section~\ref{sec:flux_tunneling}. Moreover, we reproduce our result of Section~\ref{sec:flux_tunneling} for the decay rate via a brane instanton computation. In Section~\ref{sec:background_geometry} we turn to the back-reaction of the branes on the vacuum. Specifically, in Section~\ref{sec:core} we analyze the linearized equations of motion near the $\ads$ core and compare the resulting behaviour of the fields to the corresponding ones for $\text{D}3$-branes in Type IIB string theory and for the four-dimensional Reissner-Nordstr\"om black hole. The latter represents a particularly instructive model, where one can identify the physical origin of singular perturbations. In Section~\ref{sec:infinity} we study the geometry away from the branes, solving the asymptotic equations of motion. We find a singular ``pinch-off'' at a finite (transverse) distance from the branes, as in\footnote{Indeed, our results suggest that the solution of~\cite{Dudas:2000ff} corresponds to $\text{D}8$-branes.}~\cite{Dudas:2000ff}, which hints at the idea that, in the presence of dilaton tadpoles, any breaking of ten-dimensional Poincar\'e invariance is accompanied by a finite-distance pinch-off determined by the residual symmetry. Physically, this corresponds to the fact that branes are not isolated objects in these models, since non-supersymmetric orientifolds bring along additional (anti-)$\text{D}$-branes that interact with them. Finally, in Section~\ref{sec:holographic_picture} we discuss some ramifications of our results in a holographic context. In particular, we focus on the possibility of realizing the correspondence between vacuum bubbles and RG flows that was put forth in~\cite{Antonelli:2018qwz}. We conclude in Section~\ref{sec:conclusions} with a summary of our work, some comments on its potential implications and a discussion of possible future developments.
    
\section{Anti de-Sitter vacua}\label{sec:ads_vacua}

In this section we present the (super)gravity theories related to the string vacua at stake and their $\adsts$ solutions. For the sake of generality, we shall often work with a family of $D$-dimensional effective gravitational theories, where the bosonic fields include a dilaton $\phi$ and a $(p+2)$-form field strength\footnote{The massless spectrum of the corresponding string models also includes Yang-Mills fields, but we shall not consider them. However, $\ads$ vacua supported by non-abelian gauge fields do exist~\cite{Mourad:2016xbk}.} $H_{p+2} = dB_{p+1}$. Using the ``mostly-plus'' metric signature, the (Einstein-frame) effective actions
%
\begin{equation}\label{eq:action}
    S = \frac{1}{2\kappa_D^2}\int d^D x \, \sqrt{-g} \, \left( R - \frac{4}{D-2}\, (\partial \phi)^2 - V(\phi) - \frac{f(\phi)}{2k!}\, H_{p+2}^2 \right)
\end{equation}
%
encompass all relevant cases, and furthermore we specialize them to the choices
\begin{equation}\label{eq:potential_form-coupling}
    V(\phi) = T \, e^{\gamma \phi} \, , \qquad f(\phi) = e^{\alpha \phi} \, ,
\end{equation}
which capture the lowest-order contributions in the string coupling for positive\footnote{The case $\gamma = 0$, which at any rate does not arise in string perturbation theory, would not need fluxes to stabilize the dilaton.} $\gamma$ and $T$. In the orientifold models, the dilaton potential arises from the non-vanishing NS-NS tadpole at (projective-)disk level, while in the heterotic model it arises from the torus amplitude. 
The (bosonic) low-energy dynamics of both the $USp(32)$ BSB model and the $U(32)$ Type $0'\text{B}$ model is encoded in the Einstein-frame parameters
\begin{equation}\label{eq:bsb_electric_params}
    D = 10 \, , \quad p = 1 \, , \quad \gamma = \frac{3}{2} \, , \quad \alpha = 1 \, ,
\end{equation}
whose string-frame counterpart stems from the effective action\footnote{In eq.~\eqref{eq:string_frame_action_bsb} we used the notation $F_3 = dC_2$ in order to stress the Ramond-Ramond (RR) origin of the field strength.}~\cite{Dudas:2000nv}
\begin{equation}\label{eq:string_frame_action_bsb}
    S_{\text{orientifold}} = \frac{1}{2\kappa_{10}^2}\int d^{10} x \, \sqrt{-g} \, \left( e^{-2\phi} \left[ R + 4 \, (\partial \phi)^2 \right] - T \, e^{-\phi} - \frac{1}{12}\, F_{3}^2 \right) \, .
\end{equation}
The $e^{-\phi}$ factor echoes the (projective-)disk origin of the exponential potential for the dilaton, and the coefficient $T$ is given by
\begin{equation}\label{eq:T_bsb}
    T = 2\kappa_{10}^2 \, \times 64 \, T_{\text{D}9} = \frac{16}{\pi^2 \, \alpha'}
\end{equation}
in the BSB model, and reflects the cumulative contribution of $16$ $\overline{\text{D}9}$-branes and the orientifold plane~\cite{Sugimoto:1999tx}, while in the Type $0'\text{B}$ model $T$ is half of this value.




On the other hand, the $SO(16) \times SO(16)$ heterotic model of~\cite{AlvarezGaume:1986jb} is described by
\begin{equation}\label{eq:het_electric_params}
    D = 10 \, , \quad p = 1 \, , \quad \gamma = \frac{5}{2} \, , \quad \alpha = -1 \, ,
\end{equation}
%
corresponding to the string-frame effective action
\begin{equation}\label{eq:string_frame_action_het}
    S_{\text{heterotic}} = \frac{1}{2\kappa_{10}^2}\int d^{10} x \, \sqrt{-g} \, \left( e^{-2\phi} \left[ R + 4 \, (\partial \phi)^2 - \frac{1}{12}\, H_{3}^2\right] - T  \right) \, ,
\end{equation}
which contains the Kalb-Ramond field strength $H_3$ and the one-loop cosmological constant $T$, which was estimated in~\cite{AlvarezGaume:1986jb}. One can equivalently dualize the Kalb-Ramond form and work with the Einstein-frame parameters
\begin{equation}\label{eq:het_magnetic_params}
    D = 10 \, , \quad p = 5 \, , \quad \gamma = \frac{5}{2} \, , \quad \alpha = 1 \, .
\end{equation}
%

Let us now discuss in detail the $\ads_{p+2} \times \ess^q$ flux vacua that we shall consider, which extend the ones found in~\cite{Mourad:2016xbk} to generic dimensions and form ranks. The equations of motion stemming from the action in eq.~\eqref{eq:action} are
\begin{eqaed}\label{eq:eoms}
    R_{MN} & = \tilde{T}_{MN} \, , \\
    \Box \, \phi - V'(\phi) - \frac{f'(\phi)}{2(p+2)!} \, H_{p+2}^2 & = 0 \, , \\
    d \star (f(\phi) \, H_{p+2}) & = 0 \, ,
\end{eqaed}
where the trace-reversed stress-energy tensor is
\begin{eqaed}\label{eq:stress_tensor}
    \tilde{T}_{MN} & = \frac{4}{D-2} \, \partial_M \phi \partial_N \phi + \frac{f(\phi)}{2(p+1)!} \, (H_{p+2}^2)_{MN} \\
    & + \frac{g_{MN}}{D-2} \left( V - \frac{p+1}{2(p+2)!} \, f(\phi) \, H_{p+2}^2 \right) \, .
\end{eqaed}
The $\ads_{p+2} \times \ess^q$ solution\footnote{Actually, the same solution describes a compactification on \textit{any} Einstein manifold with Ricci scalar curvature $\frac{q(q-1)}{R^2}$. This can have some bearing on perturbative stability.} takes the form
\begin{eqaed}\label{eq:adsxs_ansatz}
    ds^2 & = L^2 \, ds_{\ads_{p+2}}^2 + R^2 \, d\Omega_q^2 \, , \\
    H_{p+2} & = c \, \Vol_{\ads_{p+2}} \, , \\
    \phi & = \phi_0 \, ,
\end{eqaed}
where $ds_{\ads_{p+2}}^2$ is the unit-radius $\ads$ metric and $\Vol_{\ads_{p+2}}$ denotes the canonical volume form on $\ads_{p+2}$ with radius $R$. The dilaton is stabilized to a \textit{constant} value by the electric form flux on the sphere\footnote{The flux $n$ in eq.~\eqref{eq:electric_flux} is normalized for later convenience, albeit it is not an integer.},
\begin{equation}\label{eq:electric_flux}
    n = \frac{1}{\Omega_q}\int_{\ess^q} f \star H_{p+2} = c \, f \, R^q \, ,
\end{equation}
whose presence balances the runaway effects of the dilaton potential. Here $\Omega_q$ denotes the volume of the unit $q$-sphere. The geometry exists if and only if 
\begin{equation}\label{eq:adsxs_existence_conditions}
    \alpha > 0 \, , \qquad q > 1 \, , \qquad (q-1)\gamma - \alpha > 0 \, ,
\end{equation}
%
and using eq.~\eqref{eq:potential_form-coupling} the values of the string coupling $g_s = e^{\phi_0}$ and the curvature radii $L \, , \, R$ are given by
%
\begin{eqaed}\label{eq:ads_s_solution}
    c & = \frac{n}{g_s^\alpha R^q} \, , \\
    g_s^{(q-1)\gamma-\alpha} & = \left(\frac{(q-1)(D-2)}{(1+\frac{\gamma}{\alpha}(p+1))T}\right)^q \,\frac{2\gamma T}{\alpha n^2} \, , \\
    R^{2\frac{(q-1)\gamma-\alpha}{\gamma}} & = \left( \frac{\alpha + (p+1) \gamma}{(q-1)(D-2)}\right)^{\frac{\alpha+\gamma}{\gamma}} \left(\frac{T}{\alpha}\right)^{\frac{\alpha}{\gamma}}\frac{n^2}{2\gamma} \, , \\
    L^2 & = R^2 \left(\frac{p+1}{q-1} \cdot \frac{(p+1)\gamma+ \alpha}{(q-1)\gamma- \alpha}\right) \equiv \frac{R^2}{A} \, .
\end{eqaed}
%
%
%
From eq.~\eqref{eq:ads_s_solution} one can observe that the ratio of the curvature radii is a constant independent on $n$ but is not necessarily unity, in contrast with the case of the supersymmetric $\ads_5 \times \ess^5$ solution of Type IIB supergravity.

These solutions exhibit a number of interesting features. To begin with, they only exist in the presence of the dilaton potential, and indeed they have no counterpart in the supersymmetric case for $p \neq 3$. Moreover, the dilaton is constant, but in contrast to the supersymmetric $\ads_5 \times \ess^5$ solution its value is not a free parameter. Instead, the solution is entirely fixed by the flux number $n$. Finally, the large-$n$ limit always corresponds to a perturbative regime where both the string coupling and the curvatures are parametrically small, thus suggesting that the solution reliably captures the dynamics of string theory for its special values of $p$ and $q$.

As a final remark, let us stress that only one sign of $\alpha$ can support a vacuum with \textit{electric} flux threading the internal manifold. However, models with the opposite sign admit vacua with \textit{magnetic} flux, which can be included in our general solution dualizing the form field, and thus also inverting the sign of $\alpha$. No $\adsts$ solutions exist if $\alpha = 0$, which is the case for example for $\text{D}3$-branes in the Type $0'\text{B}$ model.

\subsection{Orientifold models}

For later convenience, let us present the explicit solution in the case of the two orientifold models. Since $\alpha = 1$ in this case, they admit $\ads_3 \times \ess^7$ solutions with electric flux, which correspond to near-horizon geometries of $\text{D}1$-brane stacks, according to the picture that we shall discuss in Section~\ref{sec:background_geometry}. On the other hand, while $\text{D}5$-branes are also present in the perturbative spectra of these models~\cite{Dudas:2001wd}, they seem to behave differently in this respect, since no corresponding $\ads_7 \times \ess_3$ vacuum exists\footnote{This is easily seen dualizing the three-form in the orientifold action \eqref{eq:bsb_electric_params}, which inverts the sign of $\alpha$, in turn violating the condition of eq.~\eqref{eq:adsxs_existence_conditions}.}. Using the values in eq.~\eqref{eq:bsb_electric_params}, one finds
\begin{eqaed}\label{eq:bsb_solution}
    g_s & = 3 \times 2^{\frac{7}{4}} \, T^{-\frac{3}{4}} n^{-\frac{1}{4}} \, , \\
    R & = 3^{\frac{1}{4}} \times 2^{\frac{5}{16}} \, T^{\frac{1}{16}} \, n^{\frac{3}{16}} \, , \\
    L^2 & = \frac{R^2}{6} \, .
\end{eqaed}
Since every parameter in this $\ads_3 \times \ess^7$ solution is proportional to a power of $n$, one can use the scalings
\begin{equation}\label{eq:bsb_scalings}
    g_s \propto n^{-\frac{1}{4}} \, , \qquad R \propto n^{\frac{3}{16}}
\end{equation}
to quickly derive some results of the following sections.

\subsection{Heterotic model}

The case of the heterotic model is somewhat subtler, since the physical parameters of eq.~\eqref{eq:het_electric_params} only allow solutions with \textit{magnetic flux},
\begin{equation}\label{eq:het_magnetic_flux}
    n = \frac{1}{\Omega_3}\int_{\ess^3} H_3 \, .
\end{equation}
The associated microscopic picture, discussed in Section~\ref{sec:background_geometry}, would involve $\text{NS}5$-branes, while the dual electric solution, which should be associated to fundamental heterotic strings, is absent. Dualities of the strong/weak type could possibly shed light on the fate of these fundamental strings, but their current understanding in the non-supersymmetric context is limited\footnote{Despite conceptual and technical issues, non-supersymmetric dualities connecting the heterotic model to open strings have been explored in~\cite{Blum:1997cs,Blum:1997gw}.}.

The corresponding Kalb-Ramond form lives on the sphere, so that dualizing it one can recast the solution in the form of eq.~\eqref{eq:ads_s_solution}, using the values in eq.~\eqref{eq:het_magnetic_params} for the parameters. The resulting $\ads_7 \times \ess^3$ solution reads
\begin{eqaed}\label{eq:het_solution}
    g_s & = 5^{\frac{1}{4}} \, T^{-\frac{1}{2}} n^{-\frac{1}{2}} \, , \\
    R & = 5^{-\frac{5}{16}} \, T^{\frac{1}{8}} \, n^{\frac{5}{8}} \, , \\
    L^2 & = 12 \, R^2 \, ,
\end{eqaed}
so that the relevant scalings are
\begin{equation}\label{eq:het_scalings}
    g_s \propto n^{-\frac{1}{2}} \, , \qquad R \propto n^{\frac{5}{8}} \, .
\end{equation}

\section{Flux tunneling of \texorpdfstring{$\adsts$}{AdS x S} vacua}\label{sec:flux_tunneling}

Let us now move on to study non-perturbative instabilities of the $\ads$ flux vacua that we described in the preceding section. These vacua feature perturbative instabilities carrying internal angular momenta~\cite{Gubser:2001zr,Basile:2018irz}, but we shall not concern ourselves with their effects, since we shall impose unbroken spherical symmetry at the outset. Alternatively, one could replace the internal sphere with an Einstein manifold whose Laplacian spectrum does not contain unstable modes, or with an orbifold that projects them out. This can be simply achieved with an antipodal $\mathbb{Z}_2$ projection in the heterotic model, while an analogous operation in the orientifold models appears more elusive~\cite{Basile:2018irz}.

However, as we shall see in the following, even in the absence of these classical instabilities the $\ads$ vacua of Section~\ref{sec:ads_vacua} would be at best metastable, since they undergo \textit{flux tunnelings} which change the flux number $n$, in the spirit of the Brown-Teitelboim scenario~\cite{Brown:1987dd,Brown:1988kg}. In order to appreciate this, it is instructive to perform a dimensional reduction over the sphere following~\cite{BlancoPillado:2009di}, retaining the dependence on a dynamical radion field $\psi$. The ansatz
%
%
%
%
\begin{equation}\label{eq:reduction_ansatz_einstein}
    ds^2 = e^{-2\frac{q}{p} \psi(x)} \, \widetilde{ds}^2_{p+2}(x) + e^{2\psi(x)} \, R_0^2 \, d\Omega^2_q \, ,
\end{equation}
where $R_0$ is an arbitrary reference radius, is warped in order to select the $(p+2)$-dimensional Einstein frame, described by $\widetilde{ds}^2_{p+2}$. Indeed, placing the dilaton and the form field on-shell results in the dimensionally reduced action
%
%
\begin{equation}\label{eq:reduced_action}
    S_{p+2} = \frac{1}{2\kappa_{p+2}^2} \int d^{p+2} x \, \sqrt{-\tilde{g}} \, \left(\tilde{R} - 2\tilde{\Lambda}\right) \, ,
\end{equation}
where the $(p+2)$-dimensional Newton's constant is
\begin{equation}\label{eq:red_newton_constant}
    \frac{1}{\kappa_{p+2}^2} = \frac{\Omega_q R_0^q}{\kappa_D^2} \, ,
\end{equation}
while the ``physical'' cosmological constant $\Lambda = -\frac{p(p+1)}{2L^2}$, associated to the frame used in the preceding section, is related to $\tilde{\Lambda}$ according to
\begin{equation}\label{eq:tilde_lambda}
    \tilde{\Lambda} = \Lambda \, e^{-2\frac{q}{p}\psi} \, ,
\end{equation}
which is a constant when the radion is on-shell, and
\begin{equation}\label{eq:on-shell_radion}
    e^{\psi} = \frac{R}{R_0} \propto n^{\frac{\gamma}{(q-1)\gamma-\alpha}} \, .
\end{equation}
Since we are working in the $(p+2)$-dimensional Einstein frame, the corresponding vacuum energy (density)
%
\begin{eqaed}\label{eq:tilde_vacuum_energy}
    \tilde{E}_0 & = \frac{2 \tilde{\Lambda}}{2\kappa_{p+2}^2} = - \, \frac{p(p+1)\Omega_q R_0^q}{2\kappa_D^2 L^2} \left(\frac{R}{R_0}\right)^{-2 \frac{q}{p}} \\
    & \propto - \, n^{- \frac{2(D-2)}{p(q-1-\frac{\alpha}{\gamma})}}
\end{eqaed}
dictates whether $n$ increases or decreases when the decay takes place.
Therefore, the two signs present in eq.~\eqref{eq:tilde_vacuum_energy} and the requirement that the vacuum energy decreases imply that this process drives the vacua to \textit{lower} values of $n$, eventually reaching outside of the perturbative regime where the semi-classical analysis is expected to be valid.

\subsection{Semi-classical decay rate} \label{sec:semiclassical_decay_rate}

Let us now compute the decay rate for flux tunneling in the semi-classical regime. To this end, standard instanton methods~\cite{Coleman:1977py,Callan:1977pt,Coleman:1980aw} provide most needed tools, but in the present case one is confined to the thin-wall approximation, which entails a flux variation\footnote{On the other hand, the extreme case $\delta n = n$ corresponds to the production of a \textit{bubble of nothing}~\cite{Witten:1981gj}.} $\delta n \ll n$, and the tension $\tau$ of the resulting bubble cannot be computed within the formalism\footnote{It is common to identify the tension of the bubble with the ADM tension of a brane soliton solution~\cite{BlancoPillado:2009di}. In our case this presents some challenges, as we shall discuss in Section~\ref{sec:infinity}.}. However, the \textit{probe} limit, in which the bubble does not affect the radion potential due to changing $n$, can be systematically improved upon~\cite{Brown:2010bc}, correcting the tension. 

We work within the dimensionally-reduced theory in $\ads_{p+2}$, using coordinates such that the relevant instanton is described by the Euclidean metric
\begin{equation}\label{eq:cdl_metric}
    ds^2_E = d\xi^2 + \rho^2(\xi) \, d\Omega^2_{p+1} \, ,
\end{equation}
so that the Euclidean on-shell action takes the form
\begin{equation}\label{eq:on-shell_S_E}
    S_E = 2 \, \Omega_{p+1} \int d\xi \, \rho(\xi)^{p+1} \left( \tilde{E}_0 - \frac{p(p+1)}{2\kappa_{p+2}^2 \rho(\xi)^2} \right) \, ,
\end{equation}
with the vacuum energy $\tilde{E}_0$, along with the associated curvature radius $\tilde{L}$, defined piece-wise by its values inside and outside of the bubble. Then, the energy constraint
\begin{equation}\label{eq:energy_constraint_cdl}
    (\rho')^2 = 1 - \frac{2\kappa_{p+2}^2}{p(p+1)} \, \tilde{E}_0 \, \rho^2 = 1 + \frac{\rho^2}{\tilde{L}^2} \, ,
\end{equation}
which stems from the Euclidean equations of motion, allows one to change variables in eq.~\eqref{eq:on-shell_S_E}, obtaining
\begin{equation}\label{eq:on-shell_S_E_final}
    S_E = - \, \frac{2p(p+1) \, \Omega_{p+1}}{2\kappa_{p+2}^2} \int d\rho \, \rho^{p-1} \sqrt{1+  \frac{\rho^2}{\tilde{L}^2}} \, .
\end{equation}
This expression defines the exponent $B = S_{\text{inst}} - S_{\text{vac}}$ in the semi-classical formula for the decay rate (per unit volume),
\begin{equation}\label{eq:decay_rate_total}
    \frac{\Gamma}{\text{Vol}} \sim \left(\text{det} \right) \times e^{-B} \, , \qquad B = B_{\text{area}} + B_{\text{vol}} \, ,
\end{equation}
in the standard fashion. The thin-wall bubble is a $(p+1)$-sphere of radius $\tilde{\rho}$ (over which the action has to be extremized), and therefore
\begin{equation}\label{eq:surface_B}
    B_{\text{area}} \sim \tilde{\tau} \, \Omega_{p+1} \, \tilde{\rho}^{\,p+1} \, ,
\end{equation}
where the tension $\tilde{\tau} = \tau \, e^{(p+1)\frac{q}{p} \psi}$ is measured in the $(p+2)$-dimensional Einstein frame. On the other hand, in the thin-wall approximation the volume term becomes
\begin{eqaed}\label{eq:volume_B}
    B_{\text{vol}} & = \frac{2p(p+1) \, \Omega_{p+1}}{2\kappa_{p+2}^2} \int_0^{\tilde{\rho}} d\rho \, \rho^{p-1} \left[\sqrt{1+  \frac{\rho^2}{\tilde{L}_{\text{vac}}^2}} - \sqrt{1+  \frac{\rho^2}{\tilde{L}_{\text{inst}}^2}} \right] \\
    & \sim - \, \epsilon \, \widetilde{\text{Vol}}(\tilde{\rho}) \, ,
\end{eqaed}
where the spacing
\begin{eqaed}\label{eq:spacing_energy}
    \epsilon & \sim \frac{d\tilde{E}_0}{dn} \, \delta n \propto n^{- \frac{2(D-2)}{p(q-1-\frac{\alpha}{\gamma})}-1} \, \delta n
\end{eqaed}
%
and the volume $\widetilde{\text{Vol}}(\tilde{\rho})$ enclosed by the bubble is computed in the $(p+2)$-dimensional Einstein frame,
\begin{eqaed}\label{eq:volumes_relation}
    \widetilde{\text{Vol}}(\tilde{\rho}) & = \tilde{L}^{p+2} \, \Omega_{p+1} \, \mathcal{V}\left( \frac{\tilde{\rho}}{\tilde{L}} \right) \, , \\
    \mathcal{V}(x) & \equiv \frac{x^{p+2}}{p+2} \, _2F_1\left(\frac{1}{2}, \frac{p+2}{2} ; \frac{p+4}{2};-x^2 \right) \, , \\
    x & \equiv \frac{\tilde{\rho}}{\tilde{L}} \, .
\end{eqaed}
All in all, the thin-wall exponent\footnote{Notice that eq.~\eqref{eq:total_B-factor} takes the form of an effective action for a $(p+1)$-brane in $\ads$ electrically coupled to $H_{p+2}$.}
\begin{equation}\label{eq:total_B-factor}
    B \sim \tau \, \Omega_{p+1} \, L^{p+1} \left[ x^{p+1} - (p+1)\beta \, \mathcal{V}(x) \right] \, , \qquad \beta \equiv \frac{\epsilon \, \tilde{L}}{(p+1) \tilde{\tau}}
\end{equation}
attains a local maximum at $x = \frac{1}{\sqrt{\beta^2 - 1}}$ for $\beta > 1$. On the other hand, for $\beta \leq 1$ the exponent is unbounded, since $B\rightarrow \infty$ as $x\rightarrow \infty$, and thus the decay rate is completely suppressed. Hence, it is crucial to study the large-flux scaling of $\beta$, which plays a role akin to an extremality parameter for the bubble. In particular, if $\beta$ scales with a negative power of $n$ nucleation is suppressed, whereas if it scales with a positive power of $n$ the extremized exponent $B$ approaches zero, thus invalidating the semi-classical computation. Therefore, the only scenario in which nucleation is both allowed and semi-classical at large $n$ is when $\beta > 1$ and is flux-independent. Physically, the bubble is super-extremal and has an $n$-independent charge-to-tension ratio. Since
\begin{equation}\label{eq:beta_cdl_value}
    \beta = v_0 \, \frac{\Omega_q \, \delta n}{2\kappa^2_{D} \tau} \, g_s^{- \frac{\alpha}{2}}  \, ,
\end{equation}
where the flux-independent constant
\begin{equation}\label{eq:v0_parameter}
    v_0 \equiv \sqrt{\frac{2(D-2)\gamma}{(p+1) ((q-1)\gamma-\alpha)}} \, ,
\end{equation}
this implies the scaling
\begin{equation}\label{eq:tau_scaling}
    \tau = T \, g_s^{- \frac{\alpha}{2}} \, ,
\end{equation}
with $T$ flux-independent and $\alpha$ as in eq.~\eqref{eq:potential_form-coupling}. In the next sections we shall verify that this is precisely the scaling expected from $\text{D}p$-branes and $\text{NS}5$-branes. To this end, we now proceed to describe a microscopic brane picture, studying probe branes in the $\adsts$ geometry and matching the semi-classical decay rate to a (Euclidean) world-volume action.

\section{Brane picture}\label{sec:brane_picture}

%
%


In this section we move the first steps towards our microscopic description of the $\ads$ vacua in terms of near-horizon geometries generated by brane stacks. While a more complete description would involve non-abelian world-volume actions coupled to the complicated dynamics driven by the dilaton potential, one can start from the simpler setting of brane instantons and probe branes moving in the $\adsts$ geometry. This allows one to retain computational control in the large-$n$ limit, while partially capturing the unstable dynamics at play. When framed in this fashion, instabilities suggest that the non-supersymmetric models at stake are generically driven to time-dependent configurations\footnote{Cosmological solutions of non-supersymmetric models indeed display interesting features~\cite{Sagnotti:2015asa,Gruppuso:2015xqa,Gruppuso:2017nap,Mourad:2017rrl,Basile:2018irz}.}, in the spirit of the considerations of~\cite{Basile:2018irz}.

We begin our analysis considering the dynamics of a $p$-brane moving in the $\ads_{p+2} \times \ess^q$ geometry of eq.~\eqref{eq:ads_s_solution}. In order to make contact with $\text{D}$-branes in orientifold models and $\text{NS}5$-branes in the heterotic model, let us consider a generic string-frame world-volume action of the form
\begin{equation}\label{eq:world-volume_action}
    S_p = - T_p \int d^{p+1} \zeta \, \sqrt{-j^* g_S} \, e^{- \sigma \phi} + \mu_p \int B_{p+1} \, ,
\end{equation}
specified by an embedding $j$ of the brane in space-time, which translates into the $D$-dimensional and $(p+2)$-dimensional Einstein-frame expressions
\begin{eqaed}\label{eq:einstein-frame_actions_brane}
    S_p & = - T_p \int d^{p+1} \zeta \, \sqrt{-j^* g} \, e^{\left(\frac{2(p+1)}{D-2} - \sigma \right) \phi} + \mu_p \int B_{p+1} \\
    & = - T_p \int d^{p+1} \zeta \, \sqrt{-j^* \tilde{g}} \, e^{\left(\frac{2(p+1)}{D-2} - \sigma \right) \phi - (p+1)\frac{q}{p}\psi} + \mu_p \int B_{p+1} \, .
\end{eqaed}
Since the dilaton is constant in the $\ads$ vacua, from eq.~\eqref{eq:einstein-frame_actions_brane} one can read off the effective tension
\begin{equation}\label{eq:effective_tension}
    \tau_p = T_p \, g_s^{\frac{2(p+1)}{D-2} - \sigma} \, .
\end{equation}
%
While in this action $T_p$ and $\mu_p$ are independent of the background, for the sake of generality we shall not assume that in non-supersymmetric models $T_p = \mu_p$.

\subsection{Brane instantons}\label{sec:brane_instantons}

In this section we reproduce the decay rate that we obtained in Section~\ref{sec:flux_tunneling} with a brane instanton computation\footnote{For more details, we refer the reader to~\cite{Brown:1987dd,Brown:1988kg,Maldacena:1998uz,Seiberg:1999xz}.}. Since flux tunneling preserves the symmetry of the internal manifold, the Euclidean branes are uniformly distributed over it, and are spherical in the Wick-rotated $\ads$ geometry. The Euclidean action for a $p$-brane specified by eq.~\eqref{eq:einstein-frame_actions_brane}, written in the $D$-dimensional Einstein frame, then reads
\begin{eqaed}\label{eq:brane_instanton_action}
    S_p^E & = \tau_p \, \text{Area} - \mu_p \, c \, \text{Vol} \\
    & = \tau_p \, \Omega_{p+1} \, L^{p+1} \left[ x^{p+1} - (p+1) \, \beta_p \, \mathcal{V}(x) \right] \, ,
\end{eqaed}
where $v_0$ is defined in eq.~\eqref{eq:v0_parameter}, and
\begin{equation}\label{eq:beta_p_def}
    \beta_p \equiv v_0 \, \frac{\mu_p}{T_p} \, g_s^{\sigma - \frac{2(p+1)}{D-2} - \frac{\alpha}{2}} \,.
\end{equation}
This result matches in form the thin-wall expression in eq.~\eqref{eq:total_B-factor}, up to the identifications of the tensions $\tau$, $\tau_p$ and the parameters $\beta$, $\beta_p$.
%
%

As we argued in the preceding section, it is reasonable to assume that $\beta_p$ does not scale with the flux, which fixes the exponent $\sigma$ to
\begin{equation}\label{eq:sigma_value}
    \sigma = \frac{2(p+1)}{D-2} + \frac{\alpha}{2} \, .
\end{equation}
This is the value that we shall use in the following. Notice that for $\text{D}p$-branes in ten dimensions, where $\alpha=\frac{3-p}{2}$, this choice gives the correct result $\sigma = 1$, in particular for $\text{D}1$-branes in orientifold models, according to eq.~\eqref{eq:bsb_electric_params}. Similarly, for $\text{NS}5$-branes in ten dimensions, the parameters in eq.~\eqref{eq:het_magnetic_params} also give the correct result $\sigma = 2$. This pattern persists even for the more ``exotic'' branes of~\cite{Bergshoeff:2005ac,Bergshoeff:2006gs,Bergshoeff:2011zk,Bergshoeff:2012jb,Bergshoeff:2015cba}, and it would be interesting to explore this direction further. Notice that in terms of the string-frame value $\alpha_S$, 
%
%
eq.~\eqref{eq:sigma_value} takes the simple form
\begin{equation}\label{eq:sigma_string-frame}
    \sigma = 1 + \frac{\alpha_S}{2}\,.
\end{equation}
Moreover, from eqs.~\eqref{eq:effective_tension} and~\eqref{eq:sigma_value} one finds that
\begin{equation}\label{eq:tau_p_scaling_sigma}
    \tau_p = T_p \, g_s^{- \frac{\alpha}{2}} 
\end{equation}
scales with the flux with the same power as $\tau$, as can be seen from eq.~\eqref{eq:tau_scaling}. Since the flux dependence of the decay rates computed extremizing eqs.~\eqref{eq:total_B-factor} and~\eqref{eq:brane_instanton_action} is determined by the respective tensions $\tau$ and $\tau_p$, they also scale with the same power of $n$. Together with eq.~\eqref{eq:sigma_value}, this provides evidence for the fact that, in the present setting, vacuum bubbles can be identified with microscopic branes, $\text{D}p$-branes in orientifold models and $\text{NS}5$-branes in the heterotic model.


Requiring furthermore that the decay rates computed extremizing eqs.~\eqref{eq:total_B-factor} and~\eqref{eq:brane_instanton_action} coincide, one is led to $\beta = \beta_p$, which implies
\begin{equation}\label{eq:beta_match}
    \mu_p = \frac{\Omega_q \, \delta n}{2\kappa^2_{D}} = \delta \left( \frac{1}{2\kappa_D^2} \int_{\ess^q} f \star H_{p+2} \right) \, ,
\end{equation}
where $\delta$ denotes the variation across the bubble wall, as expected for electrically coupled objects.

\subsection{Decay rate}\label{sec:brane_decay_rate}

Extremizing the Euclidean action over the radius, one obtains the final result for the semi-classical tunneling exponent
\begin{eqaed}\label{eq:decay_rate_p}
    S_p^E & = T_p \, L^{p+1} \, g_s^{- \frac{\alpha}{2}} \, \Omega_q \, \mathcal{B}_p\left(v_0 \, \frac{\mu_p}{T_p}\right) \\
    & \propto n^{\frac{(p+1)\gamma+\alpha}{(q-1)\gamma-\alpha}} \, ,
\end{eqaed}
where
\begin{equation}\label{eq:Bp_function}
    \mathcal{B}_p(\beta) \equiv \frac{1}{(\beta^2 - 1)^{\frac{p+1}{2}}} - \, \frac{p+1}{2} \, \beta \, \int_0^{ \frac{1}{\beta^2 - 1}} \frac{u^{\frac{p}{2}}}{\sqrt{1+u}} \, du \, .
\end{equation}
This expression includes a complicated flux-independent pre-factor, but it always scales with a positive power of $n$, consistently with the semi-classical limit. For the sake of completeness, let us provide the explicit result for non-supersymmetric string models, where the microscopic picture goes beyond the world-volume actions of eq.~\eqref{eq:world-volume_action}. Notice that we do not assume that $\mu_p = T_p$ in the non-supersymmetric setting, for the sake of generality. However, as we have already remarked in eq.~\eqref{eq:total_B-factor}, the tunneling process is allowed also in this case. This occurs because $v_0 > 1$, and thus also $\beta > 1$, in the supersymmetry-breaking backgrounds that we consider, since using eq.~\eqref{eq:bsb_electric_params} one finds
\begin{equation}\label{eq:v0_orientifold}
    (v_0)_{\text{orientifold}} = \sqrt{\frac{3}{2}}
\end{equation}
for the orientifold models, while using eq.~\eqref{eq:het_magnetic_params} one finds
\begin{equation}\label{eq:v0_heterotic}
    (v_0)_{\text{heterotic}} = \sqrt{\frac{5}{3}}
\end{equation}
for the heterotic model, where the standard Kalb-Ramond form is dualized.

For $\text{D}1$-branes in orientifold models, using the values in eq.~\eqref{eq:bsb_electric_params} one obtains
\begin{eqaed}\label{eq:decay_rate_D1}
    S_1^E & = \frac{T_1 \, L^2}{\sqrt{g_s}} \, \Omega_7 \, \mathcal{B}_1\left(\sqrt{\frac{3}{2}} \, \frac{\mu_1}{T_1}\right) \\
    & = \frac{\pi^4}{108\sqrt{2}} \, \mathcal{B}_1\left(\sqrt{\frac{3}{2}} \, \frac{\mu_1}{T_1}\right) \, T_1 \sqrt{T} \, \sqrt{n} \, , 
\end{eqaed}
and $S_1^E \approx 0.26 \, T_1 \sqrt{T n}$ if $\mu_1=T_1$.
%

For the heterotic model, using the values in eq.~\eqref{eq:het_magnetic_params} the Euclidean action of $\text{NS}5$-branes evaluates to
\begin{eqaed}\label{eq:decay_rate_NS5}
    S_5^E & = \frac{T_5 \, L^6}{\sqrt{g_s}} \, \Omega_3 \, \mathcal{B}_5\left(\sqrt{\frac{5}{3}} \, \frac{\mu_5}{T_5}\right) \\
    & = \frac{3456 \, \pi^2}{25} \, \mathcal{B}_5\left(\sqrt{\frac{5}{3}} \, \frac{\mu_5}{T_5}\right) \,  T_5 T \, n^4 \, , 
\end{eqaed}
%
and $S^E_5 \approx 337 \, T_5 T n^4$ if $\mu_5 = T_5$. In the presence of large fluxes the tunneling instability is thus far milder in the heterotic case.

\subsection{Probe branes in \texorpdfstring{$\adsts$}{AdS x S}}\label{sec:probe_branes}

After a nucleation event mediated by instantons takes place, the dynamics is encoded in the Lorentzian evolution of the bubble. Its counterpart in the microscopic brane picture is the separation of pairs of branes and anti-branes, which should then lead to brane-flux annihilation\footnote{For a discussion of this type of effect in Calabi-Yau compactifications, see~\cite{Kachru:2002gs}.}, with the negative brane absorbed by the stack and the positive one expelled out of the $\adsts$ near-horizon throat. In order to explore this perspective, we now study probe (anti-)branes moving in the $\adsts$ geometry. To this end, it is convenient to work in Poincar\'e coordinates, where the $D$-dimensional Einstein-frame metric reads
\begin{equation}\label{eq:poincare_metric_adsxs}
    ds^2 = \frac{L^2}{z^2} \left(dz^2 + dx^2_{1,p} \right) + R^2 \, d\Omega^2_q \, , \qquad dx^2_{1,p} \equiv \eta_{\mu \nu} \, dx^\mu dx^\nu \, ,
\end{equation}
embedding the brane according to the parametrization
\begin{equation}\label{eq:brane_embedding}
    j \, : \quad x^\mu = \zeta^\mu \, , \qquad z = Z(\zeta) \, , \qquad \theta^i = \Theta^i(\zeta) \, .
\end{equation}
%
Furthermore, when the brane is placed at a specific point in the internal sphere\footnote{One can verify that this ansatz is consistent with the equations of motion for linearized perturbations.}, $\Theta^i(\zeta) = \theta_0^i$, the Wess-Zumino term gives the volume enclosed by the brane in $\ads$. As a result, the action reduces to
%
%
%
\begin{eqaed}\label{eq:brane_action_adsxs}
    S_p & = - \tau_p \int d^{p+1}\! \zeta \, \left(\frac{L}{Z}\right)^{p+1} \left\{\sqrt{ 1 + \eta^{\mu\nu} \, \partial_\mu Z \, \partial_\nu Z } - \frac{c \, L}{p+1} \, \frac{\mu_p}{\tau_p}\right\} \, ,
\end{eqaed}
so that rigid, static branes are subject to a potential
\begin{eqaed}\label{eq:brane_potential}
    V_{\text{probe}}(Z) & = \tau_p \, \left(\frac{L}{Z}\right)^{p+1} \left[1 - \frac{c \, L \, g_s^{\frac{\alpha}{2}}}{p+1} \, \frac{\mu_p}{T_p} \right] \\
    & = \tau_p \, \left(\frac{L}{Z}\right)^{p+1} \left[1 - v_0 \, \frac{\mu_p}{T_p} \right] \, .
\end{eqaed}
The potential in eq.~\eqref{eq:brane_potential} indicates how rigid probe branes are affected by the $\adsts$ geometry, depending on the value of $v_0$. In particular, if $v_0 \, \frac{\abs{\mu_p}}{T_p} > 1$ positively charged branes are driven towards small $Z$ and thus exit the $\ads$ throat, while negatively charged ones are driven in the opposite direction.

Small deformations $\delta Z$ of the brane around the rigid configuration at constant $Z$ satisfy the linearized equations of motion
\begin{equation}\label{eq:linearized_eom_deformation}
   - \partial_\mu \partial^\mu \delta Z \sim \, \frac{p+1}{Z} \left(1 - v_0 \, \frac{\mu_p}{T_p} \right) - \frac{(p+1)(p+2)}{Z^2} \left(1 - v_0 \, \frac{\mu_p}{T_p} \right) \delta Z \, ,
\end{equation}
where the constant first term on the right-hand side originates from the potential of eq.~\eqref{eq:brane_potential} and affects rigid displacements, which behave as
\begin{equation}\label{eq:zero-mode_evolution}
    \frac{\delta Z}{Z} \sim \frac{p+1}{2} \left(1 - v_0 \, \frac{\mu_p}{T_p} \right) \left(\frac{t}{Z}\right)^2
\end{equation}
for small times $\frac{t}{Z} \ll 1$. On the other hand, for non-zero modes $\delta Z \propto e^{i \mathbf{k}_0 \cdot \mathbf{x} - i \omega_0 t}$ one finds the approximate dispersion relation
\begin{equation}\label{eq:deformations_improrer_dispersion}
    \omega^2_0 = \mathbf{k}^2_0 + \frac{(p+1)(p+2)}{Z^2} \left(1 - v_0 \, \frac{\mu_p}{T_p} \right) \, ,
\end{equation}
which holds in the same limit so that $Z$ remains approximately constant. In terms of the proper, red-shifted frequency $\omega_z = \sqrt{g^{tt}} \, \omega_0$ and wave-vector $\mathbf{k}_z = \sqrt{g^{tt}} \, \mathbf{k}_0$ for deformations of $Z$ in $\ads$, \eqref{eq:deformations_improrer_dispersion} reads
\begin{equation}\label{eq:deformations_dispersion}
    \omega^2_z = \mathbf{k}^2_z + \frac{(p+1)(p+2)}{L^2} \left(1 - v_0 \, \frac{\mu_p}{T_p} \right) \, .
\end{equation}
The dispersion relation of eq.~\eqref{eq:deformations_dispersion} displays a potential long-wavelength instability towards deformations of positively charged branes, which can drive them to grow in time, depending on the values of $v_0$ and the charge-to-tension ratio $\frac{\mu_p}{T_p}$. By comparison with eqs.~\eqref{eq:brane_potential} and~\eqref{eq:zero-mode_evolution}, one can see that this ``corrugation'' instability is present if and only if the branes are also repelled by the stack.


To conclude our analysis of probe-brane dynamics in the $\adsts$ throat, let us also consider small deformations $\delta \Theta$ in the internal sphere. They evolve according to the linearized equations of motion
\begin{equation}\label{eq:sphere_deformations_eom}
    -\partial_\mu \partial^\mu \delta \Theta = 0 \, ,
\end{equation}
%
%
%
%
so that these modes are stable at the linearized level.

\subsubsection{Probe (anti-)D1-branes and (anti-)NS5-branes}\label{sec:probe_bsb_het}

In the ten-dimensional orientifold models, in which the corresponding branes are $\text{D}1$-branes, $v_0 = \sqrt{\frac{3}{2}}$, so that even extremal $\text{D}1$-branes with\footnote{Verifying the charge-tension equality in the non-supersymmetric case presents some challenges. We shall elaborate upon this issue in Section~\ref{sec:infinity}.} $\mu_p = T_p$ are crucially repelled by the stack, and are driven to exit the throat towards $Z \to 0$. On the other hand $\overline{\text{D}1}$-branes, which have negative $\mu_p$, are always driven towards $Z \to +\infty$, leading to annihilation with the stack. This dynamics is the counterpart of flux tunneling in the probe-brane framework, and eq.~\eqref{eq:v0_parameter} suggests that while the supersymmetry-breaking dilaton potential allows $\ads$ vacua of this type, it is also the ingredient that allows BPS branes to be repelled. Physically, $\text{D}1$-branes are mutually BPS, but they interact with the $\overline{\text{D}9}$-branes that fill space-time. This resonates with the fact that, as we have argued in Section~\ref{sec:semiclassical_decay_rate}, the large-$n$ limit suppresses instabilities, since in this regime the interaction with $\overline{\text{D}9}$-branes is expected to be negligible~\cite{Angelantonj:1999qg,Angelantonj:2000kh}. Furthermore, the dispersion relation of eq.~\eqref{eq:deformations_dispersion} highlights an additional instability towards long-wavelength deformations of the branes, of the order of the $\ads$ curvature radius. Similarly, in the heterotic model $v_0 = \sqrt{\frac{5}{3}}$, so that negatively charged $\text{NS}5$-branes are also attracted by the stack, while positively charged ones are repelled and unstable towards sufficiently long-wavelength deformations.

The appearance of $v_0 > 1$ in front of the charge-to-tension ratio $\frac{\mu_p}{T_p}$ is suggestive of a dressed extremality parameter, which can be thought of, \textit{e.g.}, as an effective enhancement of the charge-to-tension ratio due to both dimensional reduction and supersymmetry breaking. This behaviour resonates with considerations stemming from the Weak Gravity Conjecture~\cite{ArkaniHamed:2006dz}, since the presence of branes which are (effectively) lighter than their charge would usually imply a decay channel for extremal or near-extremal objects. While non-perturbative instabilities of non-supersymmetric $\ads$ due to brane nucleation have been thoroughly discussed in the literature~\cite{Maldacena:1998uz,Seiberg:1999xz,Ooguri:2016pdq}, we stress that in the present case this phenomenon arises from microscopic branes interacting in the absence of supersymmetry.

\section{Background geometry}\label{sec:background_geometry}

In this section we study the background geometry generated by a stack of branes in the family of models described by eq.~\eqref{eq:action}. The dilaton potential brings along considerable challenges in this respect, both conceptual and technical. To begin with, there is no maximally symmetric vacuum that could act as a background, and thus in the presence of branes there is no asymptotic infinity of this type\footnote{Even if one were to conceive of a pathological Minkowski solution with $\phi = -\infty$ as a degenerate background (for instance, by introducing a cutoff), no asymptotically flat solution with $\phi \rightarrow -\infty$ can be found.}. We find, instead, that the geometry away from the branes ``pinches off'' at a finite geodesic distance, and exhibits a curvature singularity where the dilaton $\phi \to +\infty$. This resonates with the findings of~\cite{Dudas:2000ff}, and we do reconstruct the solutions therein in the case $p = 8$. These results suggest that, due to their interactions with the dilaton potential, branes cannot be described as isolated objects in these models, reflecting the probe-brane analysis of Section~\ref{sec:probe_branes}. As a consequence, identifying a sensible background string coupling or sensible asymptotic charges, such as the brane tension, appears considerably more difficult with respect to the supersymmetric case.

Despite these challenges, one can gain some insight studying the asymptotic geometry near the branes, where the $\adsts$ throat develops, and near the outer singularity, where the geometry pinches off. In Section~\ref{sec:core} we argue that the $\adsts$ solutions of Section~\ref{sec:ads_vacua} can arise as near-horizon ``cores'' of the full geometry, investigating an attractor-like behaviour of radial perturbations, which is characteristic of extremal objects and arises after a partial fine-tuning. This feature is reflected by the presence of free parameters in the asymptotic geometry away from the branes, which we construct in Section~\ref{sec:infinity}.

\subsection{Reduced dynamical system}

We begin imposing $SO(1,p) \times SO(q)$ symmetry, so that the metric is characterized by two dynamical functions $v(r) \, , b(r)$ of a transverse radial coordinate $r$. Specifically, we consider the fully generic ansatz
\begin{eqaed}\label{eq:brane_full_ansatz}
    ds^2 & = e^{\frac{2}{p+1}v - \frac{2q}{p}b} \, dx^2_{1,p} + e^{2v-\frac{2q}{p}b} \, dr^2 +e^{2b} \, R^2_0 \, d\Omega_q^2 \, , \\
    \phi & = \phi(r) \, , \\
    H_{p+2} & = \frac{n}{f(\phi)(R_0 \, e^b)^q} \Vol_{p+2} \, , \qquad \Vol_{p+2} = e^{2v - \frac{q}{p}(p+2)b} \, d^{p+1} x\,\wedge\, dr \, ,
\end{eqaed}
where $R_0$ is an arbitrary reference radius and the form field automatically solves its field equation. This gauge choice simplifies the equations of motion, which can be recast in terms of a constrained Toda-like system~\cite{Klebanov:1998yya,Dudas:2000sn}. Indeed, substituting the ansatz of eq.~\eqref{eq:brane_full_ansatz} in eq.~\eqref{eq:eoms} and taking suitable linear combinations, the resulting system can be derived by the ``reduced'' action
%
\begin{equation}\label{eq:toda_action}
    S_{\text{red}} = \int dr \left[ \frac{4}{D-2} \, {\phi'}^2 - \frac{p}{p+1} \, {v'}^2 + \frac{q(D-2)}{p} \, {b'}^2 - U(\phi,v,b) \right] \, ,
\end{equation}
where the potential is given by
\begin{equation}\label{eq:toda_potential}
    U = - \, T \, e^{\gamma \phi + 2v - \frac{2q}{p}b} - \frac{n^2}{2R_0^{2q}} \, e^{-\alpha\phi + 2v-\frac{2q(p+1)}{p}b} + \frac{q(q-1)}{R_0^2} \, e^{2v-\frac{2(D-2)}{p}b} \, ,
\end{equation}
and the equations of motion are supplemented by the zero-energy constraint
\begin{equation}\label{eq:toda_constraint}
    \frac{4}{D-2} \, {\phi'}^2 - \frac{p}{p+1} \, {v'}^2 + \frac{q(D-2)}{p} \, {b'}^2 + U = 0 \, .
\end{equation}

\subsection{\texorpdfstring{$\adsts$}{AdS x S} as near-horizon geometry}\label{sec:core}

Let us now apply the results of the preceding section to recast the $\adsts$ solution of eq.~\eqref{eq:ads_s_solution} as a near-horizon limit of the geometry described by eqs.~\eqref{eq:toda_action} and~\eqref{eq:toda_constraint}. To begin with, one can verify that the $\adsts$ solution now takes the form\footnote{Up to the sign of $r$ and rescalings of $R_0$, this realization of $\adsts$ of given $L$, $R$ in this ansatz is emphatically unique.}
\begin{eqaed}\label{eq:ads_s_toda}
    \phi & = \phi_0 \, , \\
    e^v & = \frac{L}{p+1} \, \left(\frac{R}{R_0}\right)^{-\frac{q}{p}} \, \frac{1}{-r} \, , \\
    e^b & = \frac{R}{R_0} \, ,
\end{eqaed}
where we have chosen negative values $r < 0$. This choice places the core at $r \to -\infty$, with the horizon at $r = -\infty$, while the outer singularity lies either at $r = +\infty$ or at some finite\footnote{In either case we shall find that the geodesic distance is finite.} $r = r_0$. The metric of eq.~\eqref{eq:brane_full_ansatz} can then be recast as $\adsts$ in Poincar\'e coordinates rescaling $x$ by a constant and substituting
\begin{equation}\label{eq:toda_ads_s_diffeo}
    r \mapsto - \, \frac{z^{p+1}}{p+1} \, .
\end{equation}
In supersymmetric cases, infinitely long $\ads$ throats behave as attractors going towards the horizon from infinity, under the condition on asymptotic parameters that specifies extremality. Therefore we proceed by analogy, studying linearized radial perturbations $\delta \phi \, , \delta v \, , \delta b$ around eq.~\eqref{eq:ads_s_toda} and comparing them to cases where the full geometry is known. To this end, notice that the potential of eq.~\eqref{eq:toda_potential} is factorized,
\begin{equation}\label{eq:toda_potential_factorized}
    U = e^{2v} \, \hat{U}(\phi, b) \, ,
\end{equation}
so that $v$ perturbations do not mix with $\phi$ and $b$ perturbations at the linear level. In addition, since the background values of $\phi$ and $b$ are constant in $r$, the constraint obtained linearizing eq.~\eqref{eq:toda_constraint} involves only $v$, and reads
\begin{equation}\label{eq:toda_linearized_constraint}
    2\frac{p}{p+1} \, v' \, \delta v' = \eval{\partial_v U}_{\adsts} \, \delta v = 2 \eval{U}_{\adsts}\, \delta v \,  = \frac{2p}{(p+1) r^2} \, \delta v
\end{equation}
so that
\begin{equation}\label{eq:toda_v_mode}
    \delta v \sim \text{const.} \times (-r)^{-1} \, .
\end{equation}
Thus, the constraint of eq.~\eqref{eq:toda_v_mode} retains only one mode $\sim (-r)^{\lambda_0}$ with respect to the linearized equation of motion for $\delta v$, with exponent $\lambda_0 = -1$.

On the other hand, $\phi$ and $b$ perturbations can be studied using the canonically normalized fields
\begin{equation}\label{eq:chi_field}
    \chi \equiv \left(\sqrt{\frac{8}{D-2}} \, \delta \phi \, , \, \sqrt{\frac{2q(D-2)}{p}} \, \delta b \right) \, ,
\end{equation}
in terms of which one finds
\begin{equation}\label{eq:chi_system}
    \chi'' \sim - \frac{1}{r^2} \, H_0 \, \chi 
\end{equation}
where the Hessian
\begin{equation}\label{eq:toda_hessian}
    H_{ab} \equiv \frac{\partial^2 U}{\partial \chi_a \partial \chi_b}\bigg|_{\adsts} = \frac{(H_0)_{ab}}{r^2} \, , \qquad (H_0)_{ab} = \text{const.}
\end{equation}
The substitution $t=\log(-r)$ then reduces the system to an autonomous one,
\begin{equation}\label{eq:toda_core_autonomous}
    \left(\dv[2]{}{t} - \dv{}{t}\right)\chi = - H_0 \chi \, ,
\end{equation}
so that the modes scale as $\chi \propto (-r)^{\lambda_i}$, where the $\lambda_i$ are the eigenvalues of the block matrix
\begin{equation}\label{eq:block_matrix}
    \mqty( 1 & -H_0\\1 & 0) \, .
\end{equation}
These are, in turn, given by
\begin{eqaed}\label{eq:toda_ads_s_eigenvalues}
    \lambda^{(\pm)}_{1,2} & = \frac{1 \pm \sqrt{1 - 4 \, h_{1,2}}}{2} \, , \\
    h_{1,2} & \equiv \frac{\tr(H_0) \pm \sqrt{\tr(H_0) - 4 \det(H_0)}}{2} \, ,
\end{eqaed}
where the trace and determinant of $H_0$ are given by
\begin{eqaed}\label{eq:tr_det_H0}
    \tr(H_0) & = -\frac{\alpha  \left(\gamma \, (\alpha +\gamma ) (D-2)^2-16\right) + 16 \, \gamma \, (p+1) \, (q-1)}{8 \, (p+1) \, ((q-1)\gamma-\alpha)} \, , \\
    \det(H_0) & = \frac{\alpha \, \gamma \, (D-2)^2 ((p+1)\gamma+\alpha)}{4 \, (p+1)^2 \, ((q-1)\gamma-\alpha)} \, .
\end{eqaed}
In the case of the orientifold models, one obtains the eigenvalues
\begin{equation}\label{eq:orientifold_eigenvalues}
    \frac{1 \pm \sqrt{13}}{2} \, , \qquad \frac{1 \pm \sqrt{5}}{2} \, ,
\end{equation}
while in the heterotic model one obtains the eigenvalues
\begin{equation}\label{eq:heterotic_eigenvalues}
    \pm 2 \sqrt{\frac{2}{3}} \, , \qquad 1 \pm 2 \sqrt{\frac{2}{3}} \, .
\end{equation}
All in all, in both cases one finds three negative eigenvalues and two positive ones, signaling the presence of three attractive directions as $r \to -\infty$. The remaining unstable modes should physically correspond to deformations that break extremality, resulting in a truncation of the $\adsts$ throat, and it should be possible to remove them with a suitable tuning of the boundary conditions at the outer singularity. In the next section we argue for this interpretation of unstable modes in the throat.

\subsubsection{Comparison with known solutions}

\begin{figure}
    \centering
    \scalebox{0.8}{\input{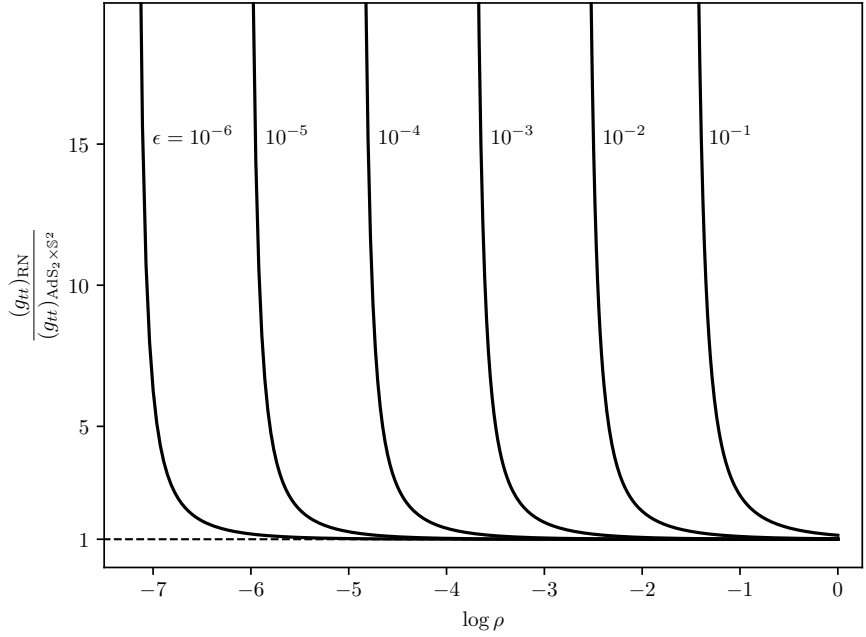}}
    \caption{A plot of the ratio of the Reissner-Nordstr\"om red-shift factor to the one of $\mathrm{AdS}_2 \times \ess^2$, for various values of the extremality parameter $\epsilon = \frac{m}{e}-1$ (only values outside of the event horizon are depicted). As extremality is approached, the horizon recedes to infinity and the geometry develops an approximate $\adsts$ throat, marked by $(g_{tt})_{\text{RN}} \approx (g_{tt})_{\ads_2 \times \ess^2}$, whose length in units of $\log \rho$ grows asymptotically linearly in $-\log \epsilon$.}
    \label{fig:throat_RN}
\end{figure}

In order to highlight the physical origin of the unstable modes, let us consider the Reissner-Nordstr\"om black hole in four dimensions, whose metric in isotropic coordinates takes the form
\begin{equation}\label{eq:RN_metric}
    ds^2_{\text{RN}} = - \frac{g(\rho)^2}{f(\rho)^2} \, dt^2 + f(\rho)^2 \, \left(d\rho^2 + \rho^2 \, d\Omega^2_2\right)\,,
\end{equation}
where
\begin{eqaed}\label{eq:RN_f_g}
    f(\rho) & \equiv 1 + \frac{m}{\rho} + \frac{m^2}{4\rho^2} - \frac{e^2}{4\rho^2} \, , \\
    g(\rho) & \equiv 1 - \frac{m^2}{4\rho^2} + \frac{e^2}{4\rho^2} \, .
\end{eqaed}
The extremal solution, $m = e$, develops an infinitely long $\ads_2 \times \ess^2$ throat in the near-horizon limit $\rho \to 0$, and radial perturbations of the type
\begin{equation}\label{eq:ads2xs2_pert}
    ds^2_{\text{pert}} = - \frac{4\rho^2}{m^2} \, e^{2 \, \delta a(\rho)} \, dt^2 + \frac{m^2}{4\rho^2} \, e^{2 \, \delta b(\rho)} \left( d\rho^2 + \rho^2 \, d\Omega^2_2\right)
\end{equation}
solve the linearized equations of motion with power-law modes $\sim \rho^{\lambda_{\text{RN}}}$, with eigenvalues
\begin{equation}\label{eq:}
    \lambda_{\text{RN}} = -2 \, , \, 1 \, , \, 0 \, .
\end{equation}
The zero-mode reflects invariance under shifts of $\delta a$, while the unstable mode reflects a breaking of extremality. Indeed, writing $m = e \, (1 + \epsilon)$ the $\frac{\rho}{m} \ll 1\, , \epsilon \ll 1$ asymptotics of the red-shift $g_{tt}$ take the schematic form
\begin{equation}\label{eq:RN_redshift}
    \frac{(g_{tt})_{\text{RN}}}{(g_{tt})_{\ads_2 \times \ess^2}} \sim \text{regular} + \epsilon \left( - \frac{1}{\rho^2} + \frac{3}{m \rho} + \text{regular} \right) + o(\epsilon) \, ,
\end{equation}
so that for $\epsilon = 0$ only a regular series in positive powers of $\rho$ remains. Geometrically, near extremality an approximate $\adsts$ throat exists for some finite length, after which it is truncated by a singularity corresponding to the event horizon. As $\epsilon$ decreases, this horizon recedes and the throat lengthens, with the length in $\log \rho$ growing as $-\log \epsilon$. This is highlighted numerically in the plot of Figure~\ref{fig:throat_RN}.

A similar analysis for BPS $\text{D}3$-branes in Type IIB supergravity~\cite{Horowitz:1991cd} yields the eigenvalues $-8 \, , \, -4 \, , \, -2 \, , \, 4 \, , \, 0 \, , 0$, suggesting again that breaking extremality generates unstable directions, and that a fine-tuning at infinity removes them leaving only the attractive ones. Notice that the zero-modes correspond to constant rescalings of $x^\mu$, which is pure gauge, and to shifts of the asymptotic value of the dilaton.

\subsection{``Pinch-off'' singularity}\label{sec:infinity}

In this section we address the asymptotic geometry away from the core. We lack a complete solution of the equations of motion stemming from eq.~\eqref{eq:toda_action}, and therefore we shall assume that the dilaton potential overwhelms the other terms of eq.~\eqref{eq:toda_potential} for large (positive) $r$, to then verify it \textit{a posteriori}. In this fashion, one can identify the asymptotic equations of motion
\begin{eqaed}\label{eq:asymptotic_toda_system}
    \phi'' & \sim \frac{\gamma(D-2)}{8} \, T \, e^{\gamma \phi + 2v - \frac{2q}{p}b} \, , \\
    v'' & \sim - \, \frac{p+1}{p} \, T \, e^{\gamma \phi + 2v - \frac{2q}{p}b} \, , \\
    b'' & \sim - \, \frac{1}{D-2} \, T \, e^{\gamma \phi + 2v - \frac{2q}{p}b} \, ,
\end{eqaed}
whose solutions
\begin{eqaed}\label{eq:asymptotic_toda_solutions}
    \phi & \sim \frac{\gamma(D-2)}{8} \, y + \phi_1 r + \phi_0 \, , \\
    v & \sim - \frac{p+1}{p} \, y + v_1 r + v_0 \, , \\
    b & \sim - \frac{1}{D-2} \, y + b_1 r + b_0
\end{eqaed}
are parametrized by the constants $\phi_{1,0} \, , v_{1,0} \, , b_{1,0}$ and a function $y(r)$ which is not asymptotically linear (without loss of generality, up to shifts in $\phi_1$, $v_1$, $b_1$). Rescaling $x$ and redefining $R_0$ in eq.~\eqref{eq:brane_full_ansatz} one can set \textit{e.g.} $b_0 = v_0 = 0$. The equations of motion and the constraint then reduce to
\begin{eqaed}\label{eq:toda_y_eom}
    y'' & \sim \hat{T} \, e^{ \Omega \, y + L \, r } \, , \\
    \frac{1}{2} \, \Omega \, {y'}^2 + L \, y' & \sim \hat{T} \, e^{ \Omega \, y + L \, r } - M \, ,
\end{eqaed}
where
\begin{eqaed}\label{eq:L_M_That}
	\hat{T} & \equiv T \, e^{\gamma \phi_0 + 2v_0 - \frac{2q}{p} \, b_0} \, , \\
	\Omega & \equiv \frac{D-2}{8} \, \gamma^2 - \frac{2(D-1)}{D-2} \, , \\
	L & \equiv \gamma \, \phi_1 + 2 \, v_1 - \frac{2q}{p} \, b_1 \, , \\
	M & \equiv \frac{4}{D-2} \, \phi_1^2 - \frac{p}{p+1} \, v_1^2 + \frac{q(D-2)}{p} \, b_1^2 \, .
\end{eqaed}
The two additional exponentials in eq.~\eqref{eq:toda_potential}, associated to flux and internal curvature contributions, are both asymptotically $\sim \exp\left( \Omega_{n,c} \, y + L_{n,c} \, r \right)$, with corresponding constant coefficients $\Omega_{n,c}$ and $L_{n,c}$. Thus, if $y$ grows super-linearly the differences $\Omega - \Omega_{n,c}$ determine whether the dilaton potential dominates the asymptotics. On the other hand, if $y$ is sub-linear the dominant balance is controlled by the differences $L - L_{n,c}$. In the ensuing discussion we shall consider the former case\footnote{The sub-linear case is controlled by the parameters $\phi_1 \, , v_1 \, , b_1$, which can be tuned as long as the constraint is satisfied. In particular, the differences $L - L_{n,c}$ do not contain $v_1$.}, since it is consistent with earlier results~\cite{Dudas:2000ff}. In order to study the system in eq.~\eqref{eq:toda_y_eom}, it is convenient to consider the cases $\Omega = 0$ and $\Omega \neq 0$ separately.

We observe that, with the choice of eq.~\eqref{eq:asymptotic_toda_solutions}, the warp exponents of the longitudinal sector $dx_{p+1}^2$ and the sphere sector $R_0 \, d\Omega_q^2$ are asymptotically equal,
\begin{equation}\label{eq:warping_asymptotic_equality}
    \frac{2}{p+1} v - \frac{2q}{p} b \sim 2b \, .
\end{equation}
This is to be expected, since if one takes a solution with $q=0$ and replaces
\begin{equation}\label{eq:longitudinal_sphere_replacement}
    dx_{p+1}^2 \rightarrow dx_{p'+1}^2 + R_0^2 \, d\Omega_{p-p'}^2
\end{equation}
for some $p'<p$ and large $R_0$, and then makes use of the freedom to rescale $R_0$ by shifting $b$ by a constant (which does not affect the leading asymptotics), one obtains another asymptotic solution with lower $p'<p$, whose warp factors are both equal to the one of the original $q = 0$ solution.

\subsubsection{Pinch-off in orientifold models}

In the orientifold models $\Omega = 0$. The system in eq.~\eqref{eq:toda_y_eom} then yields
\begin{eqaed}\label{eq:omega0_sol}
	y & \sim \frac{\hat{T}}{L^2} \, e^{L \, r} \, , \quad  M=0 \, , \qquad L > 0 \, , \\
	y & \sim \frac{\hat{T}}{2} \, r^2 \, ,\quad M= \hat{T} \,, \qquad L = 0 \, ,
\end{eqaed}
These conditions are compatible, since the quadratic form $M$ has signature $(+,-,+)$ and thus the equation $M = \hat{T} > 0$ defines a one-sheeted hyperboloid that intersects any plane, including $\{L = 0 \}$. The same is also true for the cone $\{M = 0\}$.

In both solutions the singularity arises at finite geodesic distance
\begin{equation}\label{eq:finite_distance_singularity_orientifold}
    R_c \equiv \int^{\infty} dr \, e^{v - \frac{q}{p} b} < \infty \, ,
\end{equation}
since at large $r$ the warp factor
\begin{equation}\label{eq:geodesic_radius_warping_orientifold}
    v - \frac{q}{p} \, b \sim - \, \frac{D-1}{D-2} \, y \, .
\end{equation}
In the limiting case $L = 0$, where the solution is quadratic in $r$, due to the discussion in the preceding section this asymptotic behaviour is consistent, up to the replacement of $dx_9^2$ with $dx_2^2 + R_0^2 \, d\Omega_7^2$, with the full solution found in~\cite{Dudas:2000ff}, whose singular structure is also reconstructed in our analysis for $p=8$, $q=0$, $L=0$. 

\subsubsection{Pinch-off in the heterotic model}

\begin{figure}
    \centering
    \scalebox{0.9}{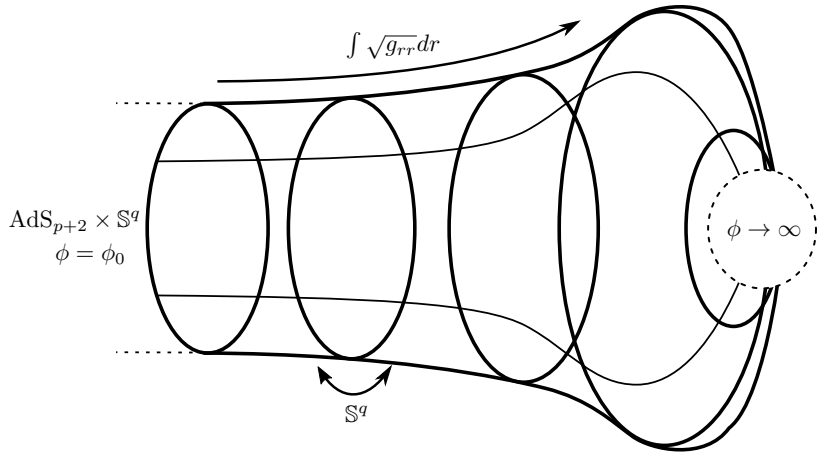}
    \caption{A schematic depiction of the expected structure of the complete geometry generated by the branes, displaying only geodesic radial distance and the $\ess^q$ radius. The geometry interpolates between the $\adsts$ throat and the pinch-off singularity (dashed circle).}
    \label{fig:geometry}
\end{figure}

In the heterotic model $\Omega = 4$, and therefore one can define
\begin{equation}\label{eq:Y_def}
	Y \equiv y + \frac{L}{\Omega} \, r\,,
\end{equation}
removing the $L \, r$ terms from the equations. One is then left with the first-order equation
\begin{equation}\label{eq:Y_eq}
	\frac{1}{2} \, {Y'}^2 - \frac{\hat{T}}{\Omega} \, e^{\Omega \, Y} = E \, ,
\end{equation}
which implies the second-order equation of motion, where the ``energy''
\begin{equation}\label{eq:energy_toda_Y}
	E \equiv \frac{M}{2\Omega} - \frac{L^2}{2\Omega^3} \, .
\end{equation}
The solutions of eq.~\eqref{eq:Y_eq} depend on the sign of $E$, and one can verify that, if $r \to +\infty$, $Y$ grows at most linearly. On the other hand, super-linear solutions develop a singularity at \textit{finite} radius $r = r_0$, and they all take the form
\begin{equation}\label{eq:Y_log_blowup}
	Y \sim - \, \frac{2}{\Omega} \, \log \left(r_0 - r \right) \, ,
\end{equation}
which is actually the exact solution of eq.~\eqref{eq:Y_eq} for $E = 0$. The geodesic distance to the singularity
\begin{equation}\label{eq:finite_distance_singularity_het}
    R_c \equiv \int^{r_0} dr \, e^{v - \frac{q}{p} b} < \infty
\end{equation}
is again finite, since from eqs.~\eqref{eq:geodesic_radius_warping_orientifold} and~\eqref{eq:Y_log_blowup}
\begin{equation}\label{eq:geodesic_radius_warping_heterotic}
    v - \frac{q}{p} \, b \sim \frac{2}{\Omega} \, \frac{D-1}{D-2} \, \log\left( r_0-r \right) = \frac{9}{16} \, \log\left( r_0-r \right) \, .
\end{equation}
In terms of the geodesic radial coordinate $\rho_c < R_c$, the asymptotics are
\begin{eqaed}\label{eq:het_DM_asymptotics}
    \phi & \sim - \, \frac{4}{5} \, \log\left( R_c - \rho_c \right) \, , \\
    ds^2 & \sim \left(R_c - \rho_c \right)^{\frac{2}{25}} \left( dx_6^2 + R_0^2 \, d\Omega_3^2 \right) + d\rho^2 \, .
\end{eqaed}
While these results are at most qualitative in this asymptotic region, they again hint at a physical picture whereby space-time pinches off at finite distance in the presence of (exponential) dilaton potentals, while branes dictate the symmetries of the geometry, as depicted in Figure~\ref{fig:geometry}. In this context, the nine-dimensional Dudas-Mourad solutions correspond to (necessarily uncharged) $8$-branes. This picture highlights the difficulties encountered in defining tension and flux as asymptotic charges, but analogous quantities might appear as parameters in the sub-leading portion of the solution, which ought to be matched with the $\adsts$ core.

As a final comment, let us add that the cosmological counterpart of these solutions, whose behaviour appears milder, can be expected to play a role when the dynamics of pinch-off singularities are taken into account.

\section{Holographic picture}\label{sec:holographic_picture}

Let us conclude our discussion with some comments and speculations on the potential holographic implications of this work. The emergence of an $\ads$ geometry in the non-supersymmetric case suggests that a dual conformal field theory (CFT) description should in principle exist, and that it ought to encode gravitational instabilities in a holographic fashion. In particular, the perturbative instabilities explored in~\cite{Gubser:2001zr,Basile:2018irz} should correspond to operators with complex anomalous dimension~\cite{Klebanov:1999um}, so that a holographic description may be able to ascertain whether their presence persists for small values of $n$. On the other hand, in~\cite{Antonelli:2018qwz} we proposed a framework to describe non-perturbative instabilities holographically in terms of RG flows, and the corresponding CFT deformations should be ``heavy'', since their effect is suppressed in the large-$n$ limit.

Starting from the brane picture that we have developed in Sections~\ref{sec:brane_picture} and~\ref{sec:background_geometry}, one can expect that the dual CFT be related to a gauge theory living on the branes' world-volume. In particular, taking $N$ $\text{D}1$-branes, so that the flux $n \propto N$, this would translate into a realization of $\ads_3$/CFT$_2$ duality in a non-supersymmetric setting. The associated central charge, determined by the Brown-Henneaux formula~\cite{Brown:1986nw}, would be
\begin{equation}\label{eq:holo-a}
    c = \frac{3L}{2G_3} = \frac{12\pi \,  \Omega_7}{\kappa^2_{10}} \, L \, R^7 \propto N^{\frac{3}{2}} \, .
\end{equation}
This grows more slowly than $N^2$, the classical number of degrees of freedom present in the gauge theory. This suggests that this two-dimensional CFT arises as a non-trivial infra-red fixed point of a world-volume gauge theory which ought to be strongly-coupled, at least at large $N$, since the effective number of degrees of freedom is parametrically smaller with respect to the classical scaling.

Within this picture, perturbative instabilities can be expected to arise from brane deformations, described by world-volume scalar fields. Moreover, the brane-flux annihilation scenario described in this paper suggests that the non-perturbative instabilities should reflect the expulsion of branes from the point of view of the stack, so that the corresponding relevant deformation~\cite{Antonelli:2018qwz} should break the gauge group according to~\cite{Witten:1998xy,Seiberg:1999xz}
\begin{eqaed}\label{eq:gauge_group_reduction}
    U(N) \qquad &\longrightarrow \qquad U(N - \delta N) \times U(\delta N) \, , \\
    USp(2N) \qquad &\longrightarrow \qquad USp(2N - 2\delta N) \times USp(2\delta N)
\end{eqaed}
in the two orientifold models\footnote{Here we assume that the gauge group be unbroken in the vacuum. If not, the breaking pattern is modified accordingly.}. Therefore, ``Higgsing'' via the separation of a small number of branes from the stack constitutes a natural candidate for the relevant deformation of the CFT, since it is not protected and may in principle grow in the infra-red. This is consistent with the considerations in~\cite{Barbon:2011zz}, where the world-volume theory of a spherical brane contains a classically marginal coupling proportional to $\frac{1}{N}$, and it gives rise to a ``Fubini instanton'' that implements the Higgsing. Characterizing precisely the relevant deformation dual to the flux tunneling process would in principle allow one to test the ``bubble/RG'' proposal of~\cite{Antonelli:2018qwz}, and more importantly it would shed some light on the behaviour of the system at small $N$, at least in the case of $\text{D}1$-branes where the dual gauge theory would be two-dimensional. We intend to pursue this possibility in a future work.

To conclude our discussion, let us mention that one could conceive compactifications on Einstein manifolds with non-trivial lower-dimensional cycles, which undergo semi-classically identical flux tunneling processes. In this case, wrapped branes could generate baryon-like Pfaffian operators~\cite{Witten:1998xy,Kachru:2002gs} in the gauge theory, which are additional candidates for relevant deformations dual to non-perturbative instabilities. However, one may anticipate that this setting could bring along subtleties due to the Myers effect~\cite{Myers:1999ps,Kachru:2002gs}.

\section{Conclusions}\label{sec:conclusions}

In this paper we have studied non-perturbative instabilities of $\ads$ vacua that arise in non-supersymmetric orientifolds, where large fluxes can provide regimes where computations are under control, and we have computed the corresponding semi-classical decay rates. To begin with, we have recast them in terms of a dismantling stack of $\text{D}1$-branes in the orientifold models, or $\text{NS}5$-branes in the heterotic model. We have provided evidence for this microscopic picture studying both the behaviour of probe branes and the geometry generated by the stack. In the former setting, we have shown that probe branes are repelled by the stack, while anti-branes are attracted to it, which hints at a brane-flux annihilation scenario via nucleation of (anti-)branes. The flux carried by the stack gradually decreases during the process, while the expelled branes constitute charged bubbles akin to the Brown-Teitelboim ones.

The nucleation of these bubbles can be equivalently described by brane instantons, and their decay rates match the results obtained in the low-energy (super)gravity. In the latter setting, we investigated the geometry induced by the back-reaction of the stack on space-time, studying the linearized field equations near the $\ads$ throat (the ``core'' region) and the asymptotic equations away from it. The field perturbations in the core region exhibit both regular modes, characteristic of extremal objects, and singular ones. The latter could be removed, in principle, by a suitable fine-tuning away from the core, which would be reminiscent of the BPS conditions on asymptotic charges of supersymmetric cases. Away from the core, in the region where the dilaton potential dominates, the asymptotic geometry exhibits singularities at finite geodesic distance. The resulting ``pinch-off'' is along the lines of the nine-dimensional solution of~\cite{Dudas:2000ff}, albeit with different symmetries, and suggests, as already stressed in~\cite{Basile:2018irz}, that non-supersymmetric settings are dynamically driven towards time-dependent configurations. In the cases that we have considered in this paper, this additional potential instability might be mitigated to an arbitrarily large extent studying the dynamics deep inside the $\ads$ throat, the deeper the more any effect of an asymptotic collapse is red-shifted.

The brane picture that we have described provides a firmer basis for further developments in brane dynamics and holography in non-supersymmetric settings. It would be interesting to build an explicit realization of the ``bubble/RG'' correspondence of~\cite{Antonelli:2018qwz} using $\text{D}1$-branes in orientifold models. More generally, our work could provide novel examples of non-supersymmetric holographic dualities, qualitatively different from $\ads_5 \times \ess^5$ orbifolds~\cite{Kachru:1998ys,Lawrence:1998ja,Bershadsky:1998mb,Bershadsky:1998cb,Schmaltz:1998bg,Erlich:1998gb}. Most importantly, correspondences of this type could provide a wider computational window to study quantum-gravitational effects on vacuum stability, potentially allowing to explore vacua with small flux numbers. In this context, computing RG flows in the dual gauge theory could provide information on the endpoint of the tunneling chain, shedding some light on the dynamics of non-supersymmetric string theory. One could conceive scenarios in which quantum effects stabilize the flux to a small value, hinting at the stringy regime as the natural one for the models at stake. Such stringy stabilization effects would conflict with widespread expectations that non-supersymmetric $\ads$ vacua are inconsistent\footnote{For a review on the status of the ``Swampland'' program, see~\cite{Palti:2019pca}.}~\cite{ArkaniHamed:2006dz,Ooguri:2016pdq,Brennan:2017rbf,Danielsson:2016rmq}. It would be interesting to explore this possibility in an explicit model, in particular in the case of $\text{D}1$-branes, since powerful analytical and numerical tools are available to study two-dimensional field theories. These and other related issues are currently under investigation.


\acknowledgments %

We are grateful to A. Sagnotti for support and suggestions, and for his feedback on the manuscript. We would like to thank A. Bombini for valuable and stimulating discussions. We also extend our gratitude to the hospitality of the Deutsches Elektronen-Synchrotron (DESY) center in Hamburg, where the present work was finalized.

The authors are supported in part by Scuola Normale Superiore and by INFN (IS CSN4-GSS-PI).


\bibliographystyle{JHEP}

\bibliography{bsb-brane.bib}

\end{document}